\documentclass[reprint,
nofootinbib,
amsmath,amssymb,
aps, prd,
]{revtex4-2}
\usepackage[utf8]{inputenc}
\usepackage{graphicx}% Include figure files
\usepackage{dcolumn}% Align table columns on the decimal point
\usepackage[colorlinks,citecolor=blue,linkcolor=blue,urlcolor=blue]{hyperref}
\usepackage{orcidlink}
\usepackage{comment}
\usepackage{tikz}
\usepackage{pgfplots}
\pgfplotsset{compat=1.18}
\usepackage{subfigure}
\usepackage{footnote}
\usepackage[title]{appendix}
\usepackage{bm}% bold math
\usepackage{xcolor}
\usepackage{verbatim} 
\begin{document}
\preprint{APS/123-QED}

\title{Complexity Growth in Black Holes: A Comparison of the Volume \\ and Action Proposals}% Force line breaks with \\

\author{Suraj Maurya\orcidlink{0000-0001-6907-8584}}
 \email{p20200471@hyderabad.bits-pilani.ac.in}%Lines break automatically or can be forced with \\

\author{Sashideep Gutti\orcidlink{0000-0001-7555-8453}}%
 \email{sashideep@hyderabad.bits-pilani.ac.in}
 
\author{Rahul Nigam\orcidlink{0000-0002-0497-5898}}%
 \email{rahul.nigam@hyderabad.bits-pilani.ac.in}

 \author{ Swastik Bhattacharya\orcidlink{0009-0008-0415-373X}}%
 \email{swastik@hyderabad.bits-pilani.ac.in}

\affiliation{Birla Institute of Technology and Science Pilani (Hyderabad Campus), Hyderabad 500078, India}
\date{\today}% It is always \today, today,

\begin{abstract}
%\begin{comment}
In this article, we investigate the late-time growth of holographic complexity, defined via the complexity--volume (CV) and complexity--action (CA) prescriptions, for BTZ, Schwarzschild, Reissner--Nordström, and Kerr black holes. Extending previous analyses beyond asymptotically AdS spacetimes, we include asymptotically flat geometries and employ the CV and CA prescriptions as comparative geometric diagnostics of black hole interior dynamics. In all cases considered, the complexity growth rate is governed by horizon thermodynamic data and scales with $T_H S_H$. While the CV prescription exhibits geometry-dependent proportionality constants, the CA prescription yields a universal thermodynamic scaling across all black holes studied, including non-AdS cases. We further analyze variations in the complexity growth rate, $\delta \dot{\mathcal{C}}$, under physical processes such as the Penrose process, superradiance, and particle accretion. We find that $\delta \dot{\mathcal{C}}$ exhibits non-trivial behavior: it increases under the Penrose process and superradiance, while under particle accretion it can increase, remain unchanged, or decrease depending on the angular momentum of the infalling particle. In quasi-equilibrium regimes, the variation in complexity closely tracks the behavior of the horizon area and interior volume growth, whereas out-of-equilibrium processes render it sensitive to angular momentum transfer and may lead to negative values within an equilibrium approximation. This behavior highlights the limitations of equilibrium-based treatments and motivates a fully dynamical analysis incorporating horizon stresses and transient hair.

%\end{comment} 
\end{abstract}

%\keywords{Suggested keywords}%Use show keys class option if keyword
\maketitle

%\tableofcontents
\section{Introduction}
\label{Section I}
Quantum complexity, originally formulated in quantum information theory as the minimal number of elementary unitary operations required to prepare a target state from a reference state, has recently emerged as a potentially powerful tool for probing black hole interiors. Motivated by the observation that certain geometric quantities inside black holes continue to grow long after thermal equilibration, it has been proposed that complexity may provide a bridge between quantum information and gravitational dynamics \cite{Susskind}. This idea has led to a growing body of work exploring geometric duals of quantum complexity within holographic frameworks. \par
Within the context of the AdS/CFT correspondence \cite{JM, EW, SS}, two principal proposals have been put forward. The complexity--volume (CV) duality relates the complexity of a boundary conformal field theory (CFT) state to the volume of a maximal spacelike hypersurface anchored at a fixed boundary time. An alternative proposal, known as the complexity--action (CA) duality, identifies complexity with the gravitational action evaluated on the Wheeler--DeWitt patch of the bulk spacetime \cite{LS2,LS3}. Both conjectures were motivated by the expectation that complexity exhibits sustained growth well beyond local thermal equilibrium, in contrast to conventional thermodynamic observables. \par
A key outcome of these developments is the expectation that, for chaotic systems in equilibrium, the late-time growth rate of complexity is controlled by macroscopic thermodynamic quantities and scales as
\begin{equation}
\frac{d\mathcal{C}}{dt} \sim \frac{T S}{\hbar},
\end{equation}
where $T$ and $S$ denote the temperature and entropy, respectively \cite{LS1,Leonard}. This scaling reflects the intuitive picture that complexity growth is governed by the number of effective degrees of freedom together with the characteristic thermal time scale. Explicit calculations in holographic CFTs have provided strong support for this relation. \par
It has been conjectured that the computational complexity of a boundary quantum state is related to the interior volume of the dual black hole spacetime. In particular, the CV duality proposes that the complexity of the boundary state is proportional to the maximal volume of a spacelike hypersurface anchored at the boundary time slice \cite{Susskind, LS1, LS2, LS3, TJ, Javier, Eliezer, Mohsen, Ming, Ming1, Bravo, RA, Mann1, Hamed}. The conjecture is expressed as
\begin{equation}\label{eqn4}
    \mathcal{C} \sim \frac{V}{\hbar G \ell},
\end{equation}
where $\ell$ denotes a characteristic geometric length scale of the spacetime and $V$ is the maximal interior volume of the spacelike hypersurface. 

For large AdS black holes this scale is set by the AdS radius, while for small black holes it is determined by the horizon radius. In the latter case the relation reduces to \cite{LS1}
\begin{equation}\label{eqn5}
    \mathcal{C} \sim \frac{V}{\hbar G r_+},
\end{equation}
where $r_+$ denotes the event horizon radius. However, such small black holes do not correspond to stable thermal equilibrium states in the dual conformal field theory, which highlights the subtlety in interpreting CV-complexity outside the regime where a well-defined holographic description is available. \par
Despite this progress, extending these ideas beyond asymptotically AdS spacetimes remains conceptually subtle. For asymptotically flat black holes, no precise boundary theory is known in which circuit complexity can be rigorously defined. However, analyses of the CV and CA prescriptions in asymptotically AdS spacetimes have been explored previously \cite{TJ}, where several conceptual subtleties were emphasized. In particular, even if one assumes the existence of a putative dual field theory, its formulation appears problematic. For rotating spacetimes such as Kerr, defining a globally well-behaved co-rotating reference frame at infinity presents additional challenges, potentially rendering the boundary description unstable or ill-defined (see the discussion in \cite{TJ}). These issues highlight that extending holographic complexity beyond AdS involves important interpretational caveats. Consequently, quantities computed using CV or CA prescriptions in such settings cannot be interpreted as complexity in a strict microscopic sense. Instead, they should be regarded as bulk geometric diagnostics inspired by holographic complexity, capturing aspects of interior evolution and response. In this work, we adopt this conservative viewpoint and focus on identifying robust, comparative features of complexity-related geometric quantities across different spacetimes and physical processes. \par
A central geometric ingredient in the CV proposal is the interior volume of a black hole. Unlike the horizon area, which admits a unique and well-established thermodynamic interpretation, black hole volume is inherently foliation-dependent. Nevertheless, it has been shown that the maximal interior volume of a black hole can be extremely large and, remarkably, grows linearly with time even for stationary geometries \cite{CR}. This observation has played a key role in motivating the conjectured connection between interior volume growth and complexity growth. \par
In our earlier work \cite{SSR3}, we revisited the computation of maximal interior volumes for rotating black holes, emphasizing that commonly employed constant-$r$ hypersurfaces do not generally correspond to true maximal slices in axially symmetric spacetimes. By determining the appropriate angle-dependent Reinhart radius for the Kerr black hole, we obtained the corresponding interior volume growth rate and studied its variation under several physically relevant processes, including the Penrose process, superradiance, particle accretion, and Hawking radiation. While several earlier works (e.g. \cite{TJ}) have investigated holographic complexity in specific black hole backgrounds, our work differs in both scope and emphasis. First, we provide a unified comparative analysis of CV and CA prescriptions across multiple geometries, including both asymptotically AdS and asymptotically flat spacetimes. Second, we go beyond equilibrium configurations by analyzing the variation of the complexity growth rate under physically relevant processes such as energy extraction and accretion. Finally, we interpret these results within a consistent framework that emphasizes the role of horizon thermodynamics and non-equilibrium dynamics, particularly in settings where no precise boundary dual is known.\par
Motivated by these considerations, the purpose of the present paper is to extend and reinterpret these results from the perspective of holographic complexity. Rather than proposing new definitions or asserting a microscopic dual description in non-AdS settings, we address a more modest but physically meaningful question: how do geometric measures associated with complexity growth behave across different black hole spacetimes, and how do they respond to energy extraction, accretion, and radiation processes? In particular, we examine to what extent the characteristic $TS$ scaling persists beyond AdS and how it is realized within the CV and CA frameworks. \par
By systematically analyzing complexity growth rates for BTZ, Schwarzschild, Reissner--Nordström, and Kerr black holes, and by comparing the behavior of CV and CA prescriptions in both equilibrium and near-equilibrium processes, we aim to clarify which features of holographic complexity are universal and which depend sensitively on spacetime geometry and dynamical effects. Our results provide a controlled setting in which the scope and limitations of holographic complexity proposals can be assessed beyond their original AdS/CFT context.
\section{Complexity Growth Rate in the CV Framework}
In this section, we examine the growth rate of holographic complexity within the CV prescription for several black hole spacetimes. Our aim is not to propose a microscopic definition of computational complexity for each geometry, but rather to test the robustness of the CV-inspired relation between the growth of interior volume and thermodynamic quantities, particularly the expected scaling with the product of temperature and entropy. Within the CV framework, the complexity growth rate is related to the rate of change of the maximal interior volume $V$ as \cite{LS1}
\begin{equation}\label{eqn8}
    \frac{d\mathcal{C}}{dt} \sim \frac{1}{\hbar G \ell}\frac{dV}{dt},
\end{equation}
where $\ell$ denotes the characteristic length scale of the spacetime. For asymptotically AdS black holes, $\ell$ is naturally identified with the AdS radius. For asymptotically flat black holes, following \cite{LS1}, the horizon radius $r_+$ is used as the relevant scale, leading to
\begin{equation}\label{eqn9}
    \frac{d\mathcal{C}}{dt} \sim \frac{1}{\hbar G r_+}\frac{dV}{dt}.
\end{equation}
Throughout this section, we work in natural units, setting $\hbar=1$, $G=1/8$ for the BTZ black hole, and $\hbar=G=1$ for four-dimensional black holes.
\subsection{BTZ Black Hole}
A BTZ black hole is a $(2+1)$-dimensional rotating solution with a negative cosmological constant \cite{BTZ}. Its metric in coordinates $(t,r,\phi)$ is
\begin{equation}\label{eqn10}
   ds^2 = -f(r) dt^2 + \frac{dr^2}{f(r)} + r^2 (f^\phi dt + d\phi)^2
\end{equation}
where the shift $f^\phi$ and lapse function $f(r)$ are defined as
\begin{equation}
f^\phi  = -\frac{J}{2r^2},\ \ f(r) = \frac{r^2}{\ell^2} - M + \frac{J^2}{4r^2}.
\end{equation}
The horizon radii $r_\pm$ are determined by $f(r_\pm)=0$,
\begin{equation}\label{eqn11}
    r^2_\pm = \frac{M\ell^2}{2}\left(1 \pm \sqrt{1 - \frac{J^2}{M^2\ell^2}}\right).
\end{equation}
The maximal interior volume is dominated by a constant $r$ hypersurface at the
Reinhart radius \cite{SSR1,SSR2}
\begin{equation}\label{eqn13}
    r_R = \ell\sqrt{\frac{M}{2}},
\end{equation}
leading to a linear growth of volume with time \cite{SSR1}
\begin{equation}\label{eqn14}
    V = 2\pi t \sqrt{M^2\ell^2 - J^2}.
\end{equation}
Substituting the value of $dV/dt$ into Eq.~(\ref{eqn8}), we obtain
\begin{equation}\label{eqn15}
     \frac{d\mathcal{C}}{dt} \sim 8\pi \times \frac{2}{\ell}\sqrt{M^2\ell^2 - J^2}.
\end{equation}
The horizon temperature and entropy of the BTZ black hole are defined as \cite{BTZ}
\begin{equation}\label{eqn16}
    T_H = \frac{M}{2\pi r_+}\sqrt{1 - \frac{J^2}{M^2\ell^2}},
    \qquad
    S_H = 4\pi r_+,
\end{equation}
yielding the product $T_HS_H$ as
\begin{equation}\label{eqn17}
    T_H S_H = \frac{2}{\ell}\sqrt{M^2\ell^2 - J^2}.
\end{equation}
Thus, from Eqs. (\ref{eqn15}) and (\ref{eqn17}), we get
\begin{equation}\label{eqn18}
     \frac{d\mathcal{C}}{dt} \sim 8\pi\, T_H S_H.
\end{equation}

\subsection{Schwarzschild Black Hole}
The metric of a Schwarzschild black hole in the coordinates $(t, r, \theta, \phi)$ is defined as
\begin{equation}\label{eqn19}
    ds^2 = -f(r)dt^2 + \frac{dr^2}{f(r)} + r^2 d\theta^2 + r^2 sin^2\theta\, d\phi^2
\end{equation}
where $f(r)=1-2M/r$. The Reinhart radius is defined as
\cite{Reinhart}
\begin{equation}\label{eqn21}
    r_R = \frac{3M}{2},
\end{equation}
and the maximal interior volume at $r_R$ is \cite{CR}
\begin{equation}\label{eqn22}
    V = 3\sqrt{3}\pi M^2 t.
\end{equation}
Substituting $r_+=2M$ and $dV/dt$ in Eq.~(\ref{eqn9}), we get
\begin{equation}\label{eqn24}
     \frac{d\mathcal{C}}{dt} \sim 3\sqrt{3}\pi \frac{M}{2}.
\end{equation}
The horizon temperature and entropy are defined as
\begin{equation}\label{eqn25}
    T_H = \frac{M}{2\pi r^2_+},\ \ S_H = \pi r^2_+
\end{equation}
and the product $T_HS_H$ becomes
\begin{equation}\label{eqn26}
    T_H S_H = \frac{M}{2}.
\end{equation}
Hence, from Eqs. (\ref{eqn24}) and (\ref{eqn26}), we get
\begin{equation}\label{eqn27}
    \frac{d\mathcal{C}}{dt} \sim 3\sqrt{3}\pi\, T_H S_H.
\end{equation}
\subsection{Reissner--Nordstr\"om Black Hole}
The metric of a Reissner--Nordström black hole in the coordinates $(t, r, \theta, \phi)$ is defined as 
\begin{equation}\label{eqn28}
    ds^2 = -f(r)dt^2 + \frac{dr^2}{f(r)} + r^2d\theta^2 + r^2 sin^2\theta\, d\phi^2
\end{equation}
where $f(r)=1-\frac{2M}{r}+\frac{Q^2}{r^2}$.The horizons and the Reinhart radius are defined as \cite{Ong, SSR3}
\begin{equation}\label{eqn30}
    r_\pm = M \pm \sqrt{M^2 - Q^2},\ \  r_R = \frac{1}{4}\left(3M + \sqrt{9M^2 - 8Q^2}\right)
\end{equation}
and the maximal interior volume at $r_R$ is \cite{Ong, SSR3}
\begin{widetext}
    \begin{equation}\label{eqn32}
    V = 4\pi t\left[\frac{1}{16}\left(3M + \sqrt{9M^2 - 8Q^2}\right)^2\left\{-Q^2 +\frac{M}{2}\left(3M + \sqrt{9M^2 - 8Q^2}\right) -\frac{1}{16}\left(3M + \sqrt{9M^2 - 8Q^2}\right)^2 \right\}\right]^{1/2}
    \end{equation}
\end{widetext}
To test the relation between the complexity growth rate and volume rate in the near extremal limit, let us define the parameter $\epsilon = \sqrt{1 - Q^2/M^2}$ and expand the horizon radius and volume rate in powers of $\epsilon$, and we get
\begin{equation}\label{eqn33}
    r_+ = M(1+\epsilon), \qquad
    \frac{dV}{dt} = 4\pi M^2\epsilon + 2\pi M^2\epsilon^3.
\end{equation}
Hence, from Eqs. (\ref{eqn9}) and (\ref{eqn33}), the complexity growth rate becomes
\begin{equation}\label{eqn34}
     \frac{d\mathcal{C}}{dt} \sim
     4\pi M\epsilon \left[1 - \epsilon + \mathcal{O}(\epsilon^2)\right].
\end{equation}
The horizon temperature and entropy are defined as
\begin{equation}\label{eqn35}
    T_H = \frac{1}{2\pi}\frac{\sqrt{M^2-Q^2}}{r_+^2+Q^2},
   \ \ S_H = \pi(r_+^2+Q^2)
\end{equation}
and the product $T_HS_H$ becomes
\begin{equation}\label{eqn36}
    T_H S_H = \frac{1}{2}\sqrt{M^2-Q^2} = \frac{M\epsilon}{2}
\end{equation}
Thus, from Eqs. (\ref{eqn34}) and (\ref{eqn36}), we get
\begin{equation}\label{eqn37}
    \frac{d\mathcal{C}}{dt} \sim
    8\pi\, T_H S_H \left[1 - \epsilon + \mathcal{O}(\epsilon^2)\right].
\end{equation}
\subsection{Kerr Black Hole}
 The restriction to spherically symmetric black holes, while technically convenient limits the scope of the analysis, as astrophysical black holes are generically rotating. The Kerr geometry therefore provides a more realistic and physically rich setting in which to examine the behaviour of holographic complexity. The absence of spherical symmetry introduces qualitatively new features including frame dragging and an ergoregion and allows one to probe genuinely dynamical processes such as the Penrose process, superradiance, particle accretion with angular momentum transfer and Hawking radiation. It is therefore important to test the CV prescription in this broader and less symmetric context. The metric of the Kerr black hole in the Boyer-Lindquist coordinates $(t, r, \theta, \phi)$ is defined as
\begin{multline}\label{eqn38}
    ds^2 = - \frac{(\Delta - a^2sin^2{\theta})}{\rho^2}dt^2- \frac{4Mra sin^2{\theta}}{\rho^2} dtd\phi + \frac{\rho^2}{\Delta}dr^2\\ + \rho^2d{\theta^2} + \frac{Asin^2{\theta}}{\rho^2}d\phi^2 
\end{multline}
The parameters, $\Delta, \rho^2, a,$ and $A$ are defined as
\begin{equation}\label{eqn39}
    \begin{split}
    \Delta = r^2 - 2Mr + a^2,\ \ \rho^2 = r^2 + a^2cos^2{\theta}\\ 
   a = J/Mc, \ \ A  = (r^2 + a^2)^2 - \Delta a^2sin^2{\theta}
    \end{split}
\end{equation}
Here, $M$ and $J$ are the spacetime's ADM mass and angular momentum. The horizons of the Kerr black hole are obtained by setting $\Delta = 0$, which are defined as
\begin{equation}\label{eqn41}
        r_- = M - \sqrt{M^2 - a^2},\ \ r_+ = M + \sqrt{M^2 - a^2}
\end{equation}
For the Kerr black hole, the maximal hypersurface is angle-dependent. In the slow-rotation limit $a/M\ll1$, the Reinhart radius is given by \cite{SSR3}
\begin{equation}\label{eqn42}
    r_R(\theta) = \frac{3M}{2} -
    \frac{a^2(14- sin^2\theta)}{36M}.
\end{equation}
The interior volume in a small $a/M$ limit grows as \cite{SSR3}
\begin{equation}\label{eqn43}
    V = 3\sqrt{3}\pi M^2 t -
    \frac{16\sqrt{3}}{9}\pi a^2 t,
\end{equation}
From Eqs. (\ref{eqn9}) and (\ref{eqn43}), we get
\begin{equation}\label{eqn45}
    \frac{d\mathcal{C}}{dt} \sim
    3\sqrt{3}\pi\frac{M}{2}
    \left[1 - 0.34\frac{a^2}{M^2}\right].
\end{equation}
The horizon temperature and entropy are defined as
\begin{equation}
    T_H = \frac{1}{2\pi}\frac{\sqrt{M^2-a^2}}{r_+^2+ a^2},
   \ \ S_H = \pi(r_+^2+a^2)
\end{equation}
and the product $T_HS_H$ becomes
\begin{equation}
    T_H S_H = \frac{1}{2}\sqrt{M^2-a^2} 
\end{equation}
and in a small $a/M$ limit, we get
\begin{equation}\label{eqn48}
    T_H S_H = \frac{M}{2}
    \left[1 - 0.50\frac{a^2}{M^2}\right],
\end{equation}
Hence, from Eqs. (\ref{eqn45}) and (\ref{eqn48}), we get
\begin{equation}\label{eqn49}
    \frac{d\mathcal{C}}{dt} \sim
    3\sqrt{3}\pi
    \left[1 + 0.16\frac{a^2}{M^2}\right] T_H S_H.
\end{equation}
\paragraph*{Interpretation and significance.}
The results obtained in this section should be interpreted as a systematic test of the robustness of the CV duality rather than as evidence for a fundamental definition of quantum computational complexity applicable to all gravitational backgrounds. In asymptotically AdS spacetimes, the CV proposal is motivated by holographic duality, where the proportionality between the complexity growth rate and the product $T_H S_H$ may be understood as a boundary reflection of thermalization and interior growth. In contrast, for asymptotically flat black holes, no established holographic boundary theory exists, and the quantity defined by Eqs.~(\ref{eqn8})--(\ref{eqn9}) should therefore be viewed as an effective geometric diagnostic, rather than a literal computational complexity in a dual field theory.\par
From this perspective, the significance of our results lies in the persistence of the scaling relation $d\mathcal{C}/dt \propto T_H S_H$ across different black hole geometries, including rotating and charged cases, together with controlled deviations governed by physical parameters such as charge and angular momentum. Within the CV duality, the proportionality coefficient in the complexity growth rate depends on the spacetime geometry and on the structure of the maximal hypersurface. This feature reflects the foliation-dependent nature of the interior volume and highlights the non-universal character of the CV duality. In particular, the maximal interior volume is defined with respect to spacelike hypersurfaces anchored at a given boundary time and therefore depends on how the interior spacetime is sliced. Since different geometries admit distinct maximal foliations, this structural dependence manifests itself through geometry-dependent proportionality constants in CV  duality growth rate. In this sense, the CV duality prescription probes the boundary-time evolution of the maximal interior hypersurface, which represents a global property of the spacetime rather than a local, covariant observable determined solely by horizon dynamics. Our analysis, therefore, reframes CV duality as a probe of spatial interior evolution and thermodynamic structure, providing a unifying but geometry-dependent characterization of black hole growth that does not rely on the existence of a known boundary dual description. It is worth noting that a complementary perspective is provided by the CA duality, which associates complexity growth with the rate of change of the Wheeler--DeWitt action and exhibits a higher degree of universality; we discuss this point in later sections.
\section{Complexity--Action (CA) duality}\label{Section V}
The CA duality provides an alternative proposal for relating quantum computational complexity to geometric quantities in gravity. In this prescription, the complexity of a quantum state is conjectured to be proportional to the on-shell gravitational action evaluated on the Wheeler--DeWitt (WDW) patch of the spacetime \cite{LS3}. Unlike the CV proposal, which associates complexity with spatial volume, the CA duality ties complexity to a covariant spacetime region bounded by null surfaces.\par
In asymptotically AdS spacetimes, the CA proposal is motivated by holography, in which the boundary theory is well-defined. In non-AdS geometries, however, no explicit boundary dual is known. In such cases, we interpret the CA prescription operationally as a gravitational diagnostic that probes the rate of change of the WDW action and allows comparison with thermodynamic quantities, rather than as a derived statement about boundary complexity. The CA conjecture is expressed as \cite{LS3}
\begin{equation}\label{eqn69}
    \mathcal{C} = \frac{\mathcal{A}}{\pi \hbar},
\end{equation}
where $\mathcal{A}$ denotes the on-shell gravitational action evaluated on the WDW patch. The total action consists of bulk and boundary contributions,
\begin{multline}\label{eqn70}
    \mathcal{A} = \frac{1}{16\pi G}\int_{\mathcal{M}} \sqrt{-g}(R-2\Lambda)\, d^4x\\
    - \frac{1}{16\pi}\int_{\mathcal{M}} \sqrt{-g} F_{\mu\nu}F^{\mu\nu}\, d^4x \\
    + \frac{1}{8\pi G}\int_{\partial\mathcal{M}} \sqrt{|h|}K\, d^3x\\
    = \mathcal{A}_{EH} + \mathcal{A}_{EM} + \mathcal{A}_{GHY},
\end{multline}
where $\mathcal{A}_{EH}$ is the Einstein--Hilbert action, $\mathcal{A}_{EM}$ the Maxwell contribution when applicable, and $\mathcal{A}_{GHY}$ the Gibbons--Hawking--York boundary term. The quantity of interest in what follows is the late-time rate of change of the action,
\begin{equation}\label{eqn71}
    \frac{d\mathcal{A}}{dt}
    = \frac{d\mathcal{A}_{EH}}{dt}
    + \frac{d\mathcal{A}_{EM}}{dt}
    + \frac{d\mathcal{A}_{GHY}}{dt}.
\end{equation}
which we discuss in the next section.
\section{Complexity growth rate in CA duality}\label{Section VI}
The complexity growth rate in the CA prescription is defined as \cite{LS2, LS3}
\begin{equation}\label{eqn72}
    \frac{d\mathcal{C}}{dt} = \frac{1}{\pi\hbar}\frac{d\mathcal{A}}{dt},
\end{equation}
and throughout this section we work in natural units with $\hbar=1$. We now evaluate this quantity explicitly for a range of black hole geometries and compare the resulting growth rates with black hole thermodynamic variables. Our emphasis is on identifying systematic patterns and deviations, rather than on proposing a microscopic interpretation in cases where a boundary dual is unknown.
\subsection{BTZ black hole}
For the rotating BTZ black hole, the Wheeler--DeWitt patch lies between the outer and inner horizons $r_+$ and $r_-$ \cite{LS3}.
The Einstein-Hilbert action is defined as
\begin{equation}\label{eqn73}
    \mathcal{A}_{EH} = \frac{1}{16\pi G}\int_{\mathcal{M}} \sqrt{-g}(R - 2\Lambda)dt dr d\phi
\end{equation}
Using the on-shell value of the Ricci scalar $R=-6/\ell^2$, the Einstein--Hilbert contribution to the action yields
\begin{equation}\label{eqn77}
    \frac{d\mathcal{A}_{EH}}{dt}
    = -\frac{2}{\ell}\sqrt{M^2\ell^2 - J^2}.
\end{equation}
The Gibbons-Hawking-York surface action is defined as
\begin{equation}\label{eqn78}
    \mathcal{A}_{GHY} = \frac{1}{8\pi G}\int_{\partial\mathcal{M}}\sqrt{|h|}K dt d\phi 
\end{equation}
where $K$ is the extrinsic curvature of constant $r$ surface defined as
\begin{equation}
    K = n^\alpha_{;\alpha} = \frac{1}{\sqrt{-g}}\frac{\partial}{\partial x^\alpha}[\sqrt{-g}\times n^\alpha]
\end{equation}
After substituting the values of $\sqrt{-g}$ and $n^\alpha$, we get
\begin{equation}\label{eqn83}
    K = \frac{1}{2r\sqrt{f(r)}}\left[rf'(r) + f(r)\right]
\end{equation}
where $f'(r) = df(r)/dr$ and at the horizons, the lapse function $f(r_\pm) = 0$. The GHY term evaluated at the horizons gives
\begin{equation}\label{eqn85}
    \frac{d\mathcal{A}_{GHY}}{dt}
    = \frac{4}{\ell}\sqrt{M^2\ell^2 - J^2},
\end{equation}
so that the total rate of change of the action becomes
\begin{equation}\label{eqn86}
    \frac{d\mathcal{A}}{dt}
    = \frac{2}{\ell}\sqrt{M^2\ell^2 - J^2}.
\end{equation}
This result agrees with earlier analyses of the BTZ geometry in the CA framework \cite{LS3}. Using Eq.~(\ref{eqn72}), the corresponding complexity growth rate is
\begin{equation}\label{eqn87}
    \frac{d\mathcal{C}}{dt}
    = \frac{1}{\pi}\frac{2}{\ell}\sqrt{M^2\ell^2 - J^2}.
\end{equation}
Expressed in terms of thermodynamic quantities, this yields
\begin{equation}\label{eqn64'}
    \frac{d\mathcal{C}}{dt}
    = \frac{1}{\pi} T_H S_H,
\end{equation}
demonstrating that in CA duality, the proportionality between complexity growth rate and $T_H S_H$ persists, albeit with a coefficient parametrically smaller than in the CV prescription.
\subsection{Schwarzschild black hole}
For the four-dimensional Schwarzschild black hole, the Wheeler--DeWitt patch extends from the horizon to the spacelike singularity \cite{LS3}. The Einstein-Hilbert action is defined as
\begin{equation}\label{eqn88}
   \mathcal{A}_{EH} = \frac{1}{16\pi G}\int_{\mathcal{M}} \sqrt{-g} R dt dr d\theta d\phi
\end{equation}
Since the spacetime is Ricci-flat $R=0$, the Einstein--Hilbert contribution vanishes,
\begin{equation}\label{eqn89}
    \frac{d\mathcal{A}_{EH}}{dt}=0.
\end{equation}
The entire contribution arises from the Gibbons--Hawking--York term, yielding
\begin{equation}\label{eqn98}
    \frac{d\mathcal{A}_{GHY}}{dt} = 2M,
\end{equation}
and therefore the total rate of change of the action becomes
\begin{equation}\label{eqn99}
    \frac{d\mathcal{A}}{dt}=2M.
\end{equation}
The resulting complexity growth rate is
\begin{equation}\label{eqn100}
    \frac{d\mathcal{C}}{dt}=\frac{2M}{\pi}.
\end{equation}
In terms of thermodynamic variables, this becomes
\begin{equation}\label{eqn73'}
    \frac{d\mathcal{C}}{dt}
    = \frac{4}{\pi} T_H S_H.
\end{equation}
Thus, even in asymptotically flat spacetime, the CA prescription yields a growth rate proportional to $T_H S_H$, supporting the interpretation of CA as a geometric probe sensitive to black hole thermodynamics.
\subsection{Reissner--Nordstr\"om black hole}
For charged black holes, the Wheeler--DeWitt patch terminates at the inner horizon rather than the singularity.  
Since the spacetime is Ricci-flat $R=0$, the Einstein--Hilbert term vanishes,
\begin{equation}
    \frac{d\mathcal{A}_{EH}}{dt}=0.
\end{equation}
The entire contribution comes from the EM and GHY action. The Einstein-Maxwell action is defined as
\begin{equation}\label{eqn101}
        \mathcal{A}_{EM} = -\frac{1}{16\pi}\int_{\mathcal{M}} \sqrt{-g}F_{\mu\nu}F^{\mu\nu}dt dr d\theta d\phi
\end{equation}
where the nonzero components of the electric field strength are
\begin{equation}\label{eqn102}
    F_{rt} = - F_{tr} = \frac{Q}{r^2}
\end{equation}
So the value of the product of $F_{\mu\nu}F^{\mu\nu}$ becomes
\begin{equation}\label{eqn103}
    \begin{split}
        F_{\mu\nu}F^{\mu\nu} = -\frac{2Q^2}{r^4}
    \end{split}
\end{equation}
Evaluating Eq. (\ref{eqn101}) using (\ref{eqn103}), the Einstein--Maxwell contribution gives
\begin{equation}\label{eqn104}
    \frac{d\mathcal{A}_{EM}}{dt}
    = \sqrt{M^2-Q^2},
\end{equation}
while the Gibbons--Hawking--York term contributes an equal amount,
\begin{equation}\label{eqn111}
    \frac{d\mathcal{A}_{GHY}}{dt}
    = \sqrt{M^2-Q^2}.
\end{equation}
The total rate of change of action is therefore
\begin{equation}\label{eqn112}
    \frac{d\mathcal{A}}{dt}
    = 2\sqrt{M^2-Q^2},
\end{equation}
leading to the complexity growth rate
\begin{equation}\label{eqn113}
    \frac{d\mathcal{C}}{dt}
    = \frac{2}{\pi}\sqrt{M^2-Q^2}.
\end{equation}
In thermodynamic form, this becomes
\begin{equation}\label{eqn84'}
    \frac{d\mathcal{C}}{dt}
    = \frac{4}{\pi} T_H S_H,
\end{equation}
again revealing proportionality to $T_H S_H$ with a reduced coefficient compared to the CV case.
\subsection{Kerr black hole}
For the Kerr black hole, the Wheeler--DeWitt patch lies between the outer and inner horizons. Since the spacetime is Ricci-flat $R=0$, the Einstein--Hilbert term vanishes,
\begin{equation}\label{eqn115}
    \frac{d\mathcal{A}_{EH}}{dt}=0.
\end{equation}
The entire contribution arises from the Gibbons--Hawking--York term, yielding
\begin{equation}\label{eqn122}
    \frac{d\mathcal{A}_{GHY}}{dt}
    = \sqrt{M^2-a^2},
\end{equation}
so that
\begin{equation}\label{eqn123}
    \frac{d\mathcal{A}}{dt}
    = \sqrt{M^2-a^2}.
\end{equation}
The corresponding complexity growth rate is
\begin{equation}\label{eqn124}
    \frac{d\mathcal{C}}{dt}
    = \frac{1}{\pi}\sqrt{M^2-a^2},
\end{equation}
which can be written as
\begin{equation}\label{eqn93'}
    \frac{d\mathcal{C}}{dt}
    = \frac{2}{\pi} T_H S_H.
\end{equation}
As in the other cases, the CA prescription produces a complexity growth rate proportional to $T_H S_H$, with a coefficient smaller than that obtained from the CV duality.
\paragraph*{Interpretation and significance.} These results suggest a clear and physically consistent interpretation of the CA prescription across different black hole geometries. In contrast to the CV duality, where complexity is tied to a particular choice of maximal spatial volume, the CA prescription associates complexity growth with the rate at which spacetime action accumulates within the Wheeler--DeWitt patch. At late times, this accumulation is dominated by contributions from regions near the horizons, where causal structure rather than spatial geometry plays the central role. The emergence of a proportionality between $d\mathcal{C}/dt$ and the product $T_H S_H$ can therefore be understood as a consequence of horizon-controlled dynamics: the entropy counts the number of effective degrees of freedom associated with the horizon, while the temperature sets the natural time scale governing the rate at which the WDW patch advances. \par
Remarkably, this scaling persists even in asymptotically flat spacetimes, where no explicit boundary dual theory is known. In such cases, the CA prescription should not be viewed as a microscopic definition of circuit complexity, but rather as a covariant gravitational diagnostic that captures how rapidly the action associated with causally accessible regions grows. The fact that the proportionality constant in CA prescription is universal and independent of the specific black hole geometry, up to simple numerical factors, stands in sharp contrast with the CV duality, where the coefficient depends sensitively on spacetime structure. This universality reflects the inherently covariant nature of the CA prescription and supports the idea that action growth provides a more robust measure of interior dynamical evolution than spatial volume alone. From this perspective, the CA results reinforce the view that complexity growth, at least at the level of gravitational diagnostics, is fundamentally governed by horizon thermodynamics rather than by details of bulk geometry.
\section{Variation in the complexity growth rate for Kerr black hole}
\label{Section IV}
In our earlier work \cite{SSR3}, we investigated the variation of the interior volume growth rate, $\delta \dot{\mathcal V}$, of a Kerr black hole under several physical processes, including the Penrose process, superradiance, particle accretion, and Hawking radiation. Those results revealed a number of nontrivial features in the behavior of $\delta \dot{\mathcal V}$. In the present section, we extend that analysis to the corresponding variation of the complexity growth rate within the CV prescription.\par
Before presenting the results, it is important to emphasize a conceptual caveat concerning the notion of black hole interior volume. In a genuinely evolving spacetime, the maximal interior volume is a teleological quantity: the maximal hypersurface anchored at a given point on the apparent horizon depends not only on data specified on a given Cauchy slice, but also on the future evolution of the spacetime, in close analogy with the event horizon. Nevertheless, much like the apparent horizon provides a quasi-local characterization of black hole boundaries, one may define an apparent interior volume using the location of the Reinhart radius on a given Cauchy slice. Throughout this section, the volume rate refers to the rate of change of this apparent interior volume, evaluated under the assumption that the black hole evolves quasi-statically and remains close to equilibrium.\par
The CV conjecture relates holographic complexity to the interior volume of black holes \cite{LS2, LS3}. Motivated by this correspondence, we examine how the variation in the complexity growth rate, $\delta \dot{\mathcal C}$ (with $\dot{\mathcal C} = d\mathcal C/dt$), responds to the physical processes considered above. From Eq.~(\ref{eqn9}), the variation in the complexity growth rate for a Kerr black hole may be written as
\begin{equation}
\label{eqn51}
\delta \dot{\mathcal C}\sim \frac{1}{r_+^2} \left(r_+ \, \delta \dot{\mathcal V} - \dot{\mathcal V} \, \delta r_+ \right),
\end{equation}
where $r_+$ denotes the event horizon radius. Without loss of generality, we take the rotation parameter $a$ (and hence the angular momentum $J$) to be positive. The Kerr horizon radius is given by
$r_+ = M + \sqrt{M^2 - a^2}$,
and in the small $a/M$ limit, its variation becomes
\begin{equation}
\label{eqn52}
\delta r_+ = 2(\delta M - \Omega_H \delta J) - 2 \Omega_H \delta J,
\end{equation}
where $\Omega_H$ is the angular velocity of the horizon. From our previous analysis \cite{SSR3}, the variation of the interior volume growth rate in this limit is
\begin{equation}
\label{eqn53}
\delta \dot{\mathcal V} = 6\sqrt{3}\pi M \left[ (\delta M - \Omega_H \delta J) - \frac{37}{27} \Omega_H \delta J \right].
\end{equation}
Substituting Eqs.~(\ref{eqn52}) and (\ref{eqn53}) into Eq.~(\ref{eqn51}) and retaining leading-order terms in $a/M$, we obtain
\begin{equation}
\label{eqn54}
\delta \dot{\mathcal C}\sim 8\left[(\delta M - \Omega_H \delta J)- 0.74\, \Omega_H \delta J \right].
\end{equation}
Details of the derivation of $\delta\mathcal{\dot C}$ are provided in Appendix~\ref{Appendix E}. We now discuss its behavior under various physical processes in the following subsections.
\subsection{Penrose process}
The Penrose process \cite{RP} provides a mechanism for extracting energy and angular momentum from the ergosphere of a Kerr black hole. In this process, the decrease in the black hole mass and angular momentum is equal to (negative of) the energy and angular momentum carried by the infalling particle \cite{Frolov, Paddy, Carrol}. A characteristic feature of the Penrose process is that the decrease in angular momentum dominates over the decrease in mass, with
$\delta M < 0$ and $\delta J \ll 0$,
subject to the inequality
\begin{equation}
\label{eqn55}
\delta M - \Omega_H \delta J > 0.
\end{equation}
Combining Eqs.~(\ref{eqn54}) and (\ref{eqn55}), we find
\begin{equation}
\label{eqn56}
\delta \dot{\mathcal C} > 0.
\end{equation}
Thus, within the CV prescription and under the quasi-equilibrium assumption, the variation in the complexity growth rate is always positive during the Penrose process.
\subsection{Superradiance}
Energy and angular momentum may also be extracted from a Kerr black hole via wave scattering in the ergosphere, a phenomenon known as superradiance \cite{RP}. For a scalar field propagating in Kerr geometry, the energy and angular momentum fluxes across the horizon are given by \cite{Paddy}
\begin{equation}
\label{eqn57}
\frac{dE}{dt} = C_1 \omega (\omega - m \Omega_H),\ \ \frac{dJ}{dt} = C_1 m (\omega - m \Omega_H),
\end{equation}
where $C_1$ is a constant and $\omega$ and $m$ are the frequency and angular momentum of the wave around the black hole spin axis \cite{Paddy}. We further obtain 
\begin{equation}
\label{eqn58}
\frac{dE}{dt} - \Omega_H \frac{dJ}{dt} = C_1 (\omega - m \Omega_H)^2
> 0.
\end{equation}
As we know that the change in the black hole's mass is equivalent to rotational energy, i.e., $dM = dE$, so from Eq. (\ref{eqn54}), we can write
\begin{equation}
\label{eqn59}
\delta \dot{\mathcal C} \sim 8\left[ \left(\frac{dE}{dt}- \Omega_H \frac{dJ}{dt} \right) - 0.74 \, \Omega_H \frac{dJ}{dt} \right] \delta t.
\end{equation}
Since both energy and angular momentum flux are radiated away from the black hole, $dE/dt < 0$ and $dJ/dt < 0$; therefore, Eqs.~(\ref{eqn58}) and (\ref{eqn59}) imply
\begin{equation}
\label{eqn60}
\delta \dot{\mathcal C} > 0.
\end{equation}
Thus the variation in the complexity growth rate increases under superradiance.
\subsection{Particle accretion}
We now consider the accretion of a particle falling from infinity into a Kerr black hole. The particle carries positive energy, $\delta M > 0$, and may have angular momentum of either sign. Classical area increase requires
\begin{equation}
\label{eqn62}
\delta M - \Omega_H \delta J > 0.
\end{equation}
 We note that for positive $\delta J$, this imposes an upper limit. Now using Eq.~(\ref{eqn54}), the variation in the complexity growth rate is
\begin{equation}
\label{eqn63}
\delta \dot{\mathcal C} \sim 8\left[(\delta M - \Omega_H \delta J) - 0.74 \, \Omega_H \delta J \right].
\end{equation}
 For $\delta J < 0$, all terms are positive, leading to
\begin{equation}
\label{eqn64}
\delta \dot{\mathcal C} > 0.
\end{equation}
If instead $\delta J > 0$, one finds
\begin{equation}
\label{eqn65}
\begin{cases}
\delta \dot{\mathcal C} > 0, & \delta J < \dfrac{\delta M}{1.74 \Omega_H},\\[6pt]
\delta \dot{\mathcal C} = 0, & \delta J = \dfrac{\delta M}{1.74 \Omega_H},\\[6pt]
\delta \dot{\mathcal C} < 0, & \delta J > \dfrac{\delta M}{1.74 \Omega_H}.
\end{cases}
\end{equation}
At first sight, the possibility of $\delta \dot{\mathcal C} < 0$ appears to be at conflict with the expectation that the complexity growth rate should increase in classical processes \cite{SH1, SH2}. However, writing $\delta \dot{\mathcal C} = T \delta S + S \delta T$, and assuming quasi-equilibrium evolution with $\delta T \approx 0$, one recovers $\delta \dot{\mathcal C} > 0$. Negative values, therefore, signal the breakdown of the equilibrium approximation. \par
More generally, when conserved charges are present, the complexity growth rate is bounded by \cite{LS3}
\begin{equation}
\label{Complexitybound}
\frac{d\mathcal C}{dt}\le \frac{2}{\pi \hbar}\left[(M - \mu \mathcal Q) - (M - \mu \mathcal Q)_{\mathrm{gs}}
\right],
\end{equation}
where the subscript ``gs" (stands for ground state) indicates the state of lowest $(M-\mu \mathcal{Q})$ for a given chemical potential $\mu$. The chemical potential is defined as
\begin{equation}\nonumber
    \mu = \begin{cases}
        \Phi_E;\ \ \mbox{for a charged black hole}\\
        \Omega_H;\ \ \mbox{for a rotating black hole}
    \end{cases}
\end{equation}
where $\Phi_E$ and $\Omega_H$ are the electrostatic potential and horizon's angular momentum of black holes, respectively. For an asymptotically flat Kerr black hole, transient hair produced during accretion is radiated away as the system relaxes to equilibrium, and deviations from the bound are expected to be short-lived. Accounting for such non-equilibrium effects, for instance, via the membrane paradigm \cite{Thorne}, may restore $\delta \dot{\mathcal C} \ge 0$ even in cases where the leading-order estimate suggests otherwise.
\subsection{Hawking radiation}
Finally, we consider Hawking radiation. During evaporation, the black hole area decreases, so $\delta A < 0$ and $\delta M - \Omega_H \delta J < 0$, while angular momentum is lost more rapidly than mass \cite{Frolov}, implying $\delta J < 0$. From Eq.~(\ref{eqn54}),
\begin{equation}
\label{eqn68}
\delta \dot{\mathcal C}
\sim
8\left[
(\delta M - \Omega_H \delta J)
-
0.74 \, \Omega_H \delta J
\right].
\end{equation}
Here, the first term is negative, whereas the second is positive, and within the small-$J$ approximation, the sign of $\delta \dot{\mathcal C}$ depends on the detailed balance between these contributions. The result is therefore inconclusive at this order, and a more refined treatment is left for future work.
\section{Conclusions}
\label{Section VIII}
In this work, we have carried out a systematic study of the growth rate of holographic complexity for a broad class of black hole spacetimes, including BTZ, Schwarzschild, Reissner--Nordström, and Kerr geometries, using both the complexity--volume (CV) and complexity--action (CA) proposals. Our analysis encompasses both asymptotically AdS and asymptotically flat spacetimes and is therefore well suited to disentangle features that are universal from those that depend sensitively on the spacetime geometry.

A central result of our study is that, across all cases considered, the late-time complexity growth rate is governed by macroscopic horizon data. In particular, we find that the growth rate is proportional to the product of the horizon temperature and entropy, $T_H S_H$, with a geometry-dependent proportionality constant in the CV prescription. This behavior highlights the non-universal nature of CV duality, in agreement with earlier observations \cite{LS1, TJ}, while simultaneously emphasizing that the dominant control parameters are thermodynamic in character and phenomenological, rather than emerging from a fundamental microscopic dual description. Importantly, this connection does not rely on the existence of a specific boundary dual and therefore remains meaningful even for asymptotically flat black holes, where a conventional holographic interpretation is absent.

For the CV duality, we explicitly derived the proportionality constants for BTZ, Schwarzschild, Reissner--Nordström, and Kerr black holes. In the rotating case, our analysis goes beyond earlier treatments by employing a more general maximal hypersurface with explicit angular dependence, which is essential for Kerr spacetime. In the small $a/M$ regime, we showed that the leading behavior of the complexity growth rate continues to scale with $T_H S_H$, with subleading corrections controlled by powers of $a/M$. A similar structure appears in the near-extremal Reissner--Nordström case, where corrections arise as a function of $Q/M$. While the precise physical interpretation of these correction terms remains open, their presence suggests that deviations from universality encode detailed information about charge and rotation.

We also examined the complexity growth rate using the CA proposal for the same set of black hole spacetimes. In contrast to the CV case, we found that CA duality yields a single universal proportionality constant relating $d\mathcal{C}/dt$ to $T_H S_H$, even in non-AdS backgrounds. This supports the view that CA duality captures a more robust and geometry-independent notion of complexity growth, consistent with earlier arguments advocating its universality \cite{LS1}. Taken together, our CV and CA results provide a clear comparative framework for assessing how different holographic prescriptions encode the dynamics of black hole interiors.

Beyond equilibrium configurations, we investigated how the variation of the complexity growth rate responds to physical processes such as the Penrose process, superradiance, particle accretion, and Hawking radiation. We found that, in quasi-equilibrium situations such as the Penrose process and superradiance, the variation $\delta \dot{\mathcal{C}}$ closely tracks the behavior of the horizon area and the growth rate of the interior volume. In contrast, particle accretion and Hawking radiation can drive the system out of equilibrium, leading to situations in which $\delta \dot{\mathcal{C}}$ becomes sensitive to the details of angular momentum transfer and, in some cases, may even appear negative within the equilibrium approximation.

We emphasized that such negative values signal the limitations of describing highly dynamical processes using equilibrium data alone. In these regimes, transient ``hair'' and horizon stresses are expected to play a crucial role, rendering the growth of complexity effectively path-dependent. This observation naturally points toward a membrane or fluid-dynamical description of the black hole horizon as an appropriate framework for understanding complexity growth away from equilibrium. A detailed treatment of these effects, requiring explicit solutions of the perturbation equations and a fully time-dependent analysis of both CV and CA dualities, is left for future work.

These observations suggest several natural directions for future investigation. A first priority is a fully time-dependent treatment of complexity growth, in which the maximal volume surfaces and Wheeler--DeWitt patches are constructed in genuinely dynamical spacetimes rather than inferred from quasi-equilibrium data. Such an analysis would clarify the role of horizon stresses, transient hair, and angular momentum transport in regulating complexity growth beyond the equilibrium regime. In this context, it would be particularly interesting to explore whether an effective membrane or fluid description of the horizon can provide a unified framework for understanding the response of both CV and CA dualities to non-equilibrium perturbations. More broadly, extending the present analysis to collapse geometries, Vaidya-type spacetimes, or settings with time-dependent couplings may help sharpen the distinction between universal thermodynamic contributions and genuinely dynamical effects in holographic complexity.

Overall, our results demonstrate that the growth of holographic complexity, when formulated geometrically, provides a meaningful and calculable probe of black hole interior dynamics across a wide range of spacetimes. Even in the absence of a precise microscopic interpretation, the robust correlation with horizon thermodynamics and the controlled deviations induced by charge, rotation, and non-equilibrium processes offer valuable insight into the universal and non-universal aspects of complexity in gravitational systems.
\section*{ACKNOWLEDGMENTS}
We thank our institute, BITS Pilani Hyderabad campus, for providing the required infrastructure for this research work. S. M. thanks the funding agency, Council of Scientific and Industrial Research (CSIR), Government of India, File No. 09/1026(11329)/2021-EMR-I, for providing the necessary fellowship to support this research work.
\appendix
\section*{Appendix: Gravitational Action in the Wheeler-DeWitt patch}
Gravitational action in the Wheeler-DeWitt patch is equal to the sum of the actions in the bulk, such as Einstein-Hilbert and Einstein-Maxwell, and at the boundary, such as the Gibbons-Hawking-York action, which is defined as
\begin{multline}\label{WDW}
    \mathcal{A} = \frac{1}{16\pi G}\int_{\mathcal{M}} \sqrt{-g} (R-2\Lambda) d^4x\\ - \frac{1}{16\pi}\int_{\mathcal{M}} \sqrt{-g} F_{\mu\nu}F^{\mu\nu}d^4x\\
    +\frac{1}{8\pi G}\int_{\partial\mathcal{M}} \sqrt{|h|}Kd^3x\\
    = \mathcal{A}_{EH} + \mathcal{A}_{EM} + \mathcal{A}_{GHY}
\end{multline}
where $g$ is the determinant of the metric tensor $g_{\mu\nu}$, $h$ is the determinant of induced metric tensor $h_{ab}$ on the constant $r$ surface, $R$ is the Ricci scalar, $\Lambda$ is the cosmological constant, $G$ is the Newton's gravitational constant, and $F_{\mu\nu}$ is the Maxwell field tensor. Hence, the total rate of change of action in the Wheeler-DeWitt patch is defined as
\begin{equation}\label{actionrate}
    \frac{d\mathcal{A}}{dt} = \frac{d\mathcal{A}_{EH}}{dt} + \frac{d\mathcal{A}_{EM}}{dt} + \frac{d\mathcal{A}_{GHY}}{dt}
\end{equation}
The rate of change of action $\mathcal{A}$ is used to calculate the
complexity growth rate.
\section{BTZ BLACK HOLE}\label{Appendix A}
\subsection{Einstein-Hilbert action}
A BTZ black hole is a (2+1) dimensional neutral rotating black hole with an AdS background. The Einstein field equation in the presence of a cosmological constant is defined as
\begin{equation}\label{A1}
    R_{\mu\nu} - \frac{1}{2}R g_{\mu\nu} +\Lambda g_{\mu\nu} = \frac{8\pi G}{c^4}T_{\mu\nu}
\end{equation}
For vacuum space $T_{\mu\nu} = 0$. Taking the trace of Eq. (\ref{A1}) on both sides, we get
\begin{equation}\label{A2}
    \begin{split}
            Tr(R_{\mu\nu}) - \frac{1}{2}R Tr(g_{\mu\nu}) + \Lambda Tr(g_{\mu\nu}) = 0\\  \Rightarrow R - \frac{3}{2}R + 3\Lambda = 0\ \  \Rightarrow R = 6\Lambda = -\frac{6}{\ell^2}
    \end{split}
\end{equation}
The Einstein-Hilbert action for a BTZ black hole is defined as
\begin{multline}\label{A3}
   \mathcal{A}_{EH} = \frac{1}{16\pi G}\int_{\mathcal{M}} \sqrt{-g}(R - 2\Lambda)d^3x\\ = -\frac{1}{2\pi}\times\frac{4}{\ell^2}\int \sqrt{-g} dt dr d\phi
\end{multline}
where we substitute $G = 1/8$ into the above expression. The Einstein-Hilbert action includes a coefficient $1/16\pi G$. By choosing $G=1/8$, the prefactor simplifies to $1/2\pi$, making analytical calculations easier and significantly simplifying the Einstein field equation. Now, the determinant of the BTZ metric is $g = -r^2$. Substituting the value of $g$ in Eq. (\ref{A3}), the rate of change of action becomes
\begin{multline}\label{A4}
    \frac{d\mathcal{A}_{EH}}{dt} = - \frac{1}{2\pi}\times\frac{4}{\ell^2}\int_{r_-}^{r_+} r dr \int_{0}^{2\pi} d\phi\\ = -\frac{1}{2\pi}\times\frac{4}{\ell^2}\times 2\pi \times \left[\frac{r^2}{2}\right]_{r_-}^{r_+} = -\frac{2}{\ell^2}\left[r^2_+ - r^2_-\right]\\ = - \frac{2}{\ell}\sqrt{M^2\ell^2 - J^2}
\end{multline}
\subsection{Gibbons-Hawking-York action}
The metric of a BTZ black hole is defined as 
\begin{multline}\label{A5}
    ds^2 =-f(r) dt^2 + \frac{dr^2}{f(r)} + r^2 (f^\phi dt + d\phi)^2\\ = -\left[f(r) - \frac{J^2}{4r^2}\right]dt^2 + \frac{dr^2}{f(r)} - Jdt d\phi + r^2d\phi^2
\end{multline}
where $f^\phi = -J/2 r^2$ and $f(r) =  \frac{r^2}{\ell^2} - M + \frac{J^2}{4 r^2}$ are known as shift function and lapse function respectively. Determinant of the metric (\ref{A5}) is defined as
\begin{equation}\label{A6}
    g = det(g_{\mu\nu}) = -r^2
\end{equation}
Here $g_{rr} = 1/f(r)$. The unit normal vector to a constant $r$ surface is defined as
\begin{equation}\label{A7}
    n^{\alpha}n_{\alpha} = 1 \Rightarrow n^r n_r = n^r g_{rr} n^r = 1 \Rightarrow n^r = \frac{1}{\sqrt{g_{rr}}} = \sqrt{f(r)}
\end{equation}
Now, the trace of the extrinsic curvature is defined as
\begin{multline}\label{A8}
    K = n^\alpha_{;\alpha} = \frac{1}{\sqrt{-g}}\frac{\partial}{\partial x^\alpha}[\sqrt{-g}\times n^\alpha]= \frac{1}{\sqrt{-g}}\frac{\partial}{\partial r}[\sqrt{-g}\times n^r]\\ = \frac{1}{r}\frac{\partial}{\partial r}\left[r\sqrt{f(r)}\right] = \frac{1}{2r\sqrt{f(r)}}\left[rf'(r) + f(r)\right]
\end{multline}
where $f'(r) = df(r)/dr$. The induced metric for a constant $r$ surface becomes
\begin{equation}\label{A9}
   ds^2 = -\left[f(r) - \frac{J^2}{4r^2}\right]dt^2 - Jdt d\phi + r^2d\phi^2
\end{equation}
Determinant of the metric (\ref{A9}) is defined as
\begin{equation}\label{A10}
    h = det (h_{ab}) = -r^2 f(r)
\end{equation}
Now, in the natural unit $\hbar = c = 1$ and $G = 1/8$, the Gibbon-Hawking-York boundary action is defined as
\begin{multline}\label{A11}
    \mathcal{A}_{GHY} = \frac{1}{8\pi G}\int_{\partial \mathcal{M}} \sqrt{|h|}K d^2x = \frac{1}{\pi}\int \sqrt{|h|}K dt d\phi\\ = \frac{1}{\pi}\int dt \int  r\sqrt{f(r)}\times\frac{1}{2r\sqrt{f(r)}}\left[rf'(r) + f(r)\right]d\phi
\end{multline}
The rate of change of the Gibbons-Hawking-York boundary action is defined as
\begin{multline}\label{A12}
    \frac{d\mathcal{A}_{GHY}}{dt} =  \frac{1}{2\pi}\left[rf'(r) + f(r)\right] \int_{0}^{2\pi} d\phi\\ =  \frac{1}{2\pi}\left[rf'(r) + f(r)\right] \times 2\pi = \left[rf'(r) + f(r)\right]
\end{multline}
At horizons where $f(r_\pm) = 0$, and with $f'(r) = \frac{df(r)}{dr} = \frac{2}{\ell^2} - \frac{J^2}{2r^3}$, the boundary of the Wheeler-DeWitt patch for a BTZ black hole extends from the outer $r_+$ to the inner horizon $r_-$. Therefore, the GHY action at $r=r_-$ and $r= r_+$ is
\begin{multline}\label{A13}
    \frac{d\mathcal{A}_{GHY}}{dt} = \left[rf'(r)\right]_{r_-}^{r_+} = \left[\frac{2r^2}{\ell^2} - \frac{J^2}{2r^2}\right]_{r_-}^{r_+}\\ = \left[\frac{2(r^2_+ - r^2_-)}{\ell^2} - \frac{J^2(r^2_+ - r^2_-)}{2r^2_+ r^2_-}\right] = \frac{4}{\ell}\sqrt{M^2\ell^2 - J^2}
\end{multline}
From Eqs. (\ref{A3}) and (\ref{A13}), the total rate of change of action is defined as
\begin{equation}\label{A14}
    \frac{d\mathcal{A}}{dt} = \frac{d\mathcal{A}_{EH}}{dt} + \frac{d\mathcal{A}_{GHY}}{dt} =  \frac{2}{\ell}\sqrt{M^2\ell^2 - J^2}
\end{equation}
Therefore, the complexity growth rate becomes
\begin{equation}\label{A15}
    \frac{d\mathcal{C}}{dt} = \frac{1}{\pi \hbar}\frac{d\mathcal{A}}{dt} = \frac{2}{\pi\ell}\sqrt{M^2\ell^2 - J^2}
\end{equation} 
where we put $\hbar = 1$ in the final expression. As we know, the horizon temperature $T_H$ and entropy $S_H$ of the BTZ black hole are defined as
\begin{multline}\label{A16}
    T_H = \frac{M}{2\pi r_+}\sqrt{1 - \frac{J^2}{M^2\ell^2}}, \ \ S_H = 4\pi r_+ \\ \Rightarrow T_H S_H = \frac{2}{\ell}\sqrt{M^2\ell^2 - J^2}
\end{multline}
Hence, from Eqs. (\ref{A15}) and (\ref{A16}), the complexity growth rate in terms of product $T_HS_H$ becomes
\begin{equation}\label{A17}
    \frac{d\mathcal{C}}{dt} = \frac{1}{\pi}T_H S_H
\end{equation}
\section{SCHWARZSCHILD BLACK HOLE}\label{Appendix B}
\subsection{Einstein-Hilbert action}
The Einstein-Hilbert action is defined as
\begin{equation}\label{B1}
   \mathcal{A}_{EH} = \frac{1}{16\pi G}\int_{\mathcal{M}} \sqrt{-g} R d^4x = \frac{1}{16\pi G}\int \sqrt{-g} R dt dr d\theta d\phi
\end{equation}
As we know, the Ricci scalar $R = 0$ for the Schwarzschild black hole in vacuum space. Hence, the rate of change of action is
\begin{equation}\label{B2}
    \frac{d \mathcal{A}_{EH}}{dt} = \frac{1}{16\pi }\int \sqrt{-g} R dr d\theta d\phi = 0
\end{equation}
\subsection{Gibbons-Hawking-York action}
The metric of a Schwarzschild black hole is defined as 
\begin{equation}\label{B3}
    ds^2= -f(r)dt^2 + \frac{dr^2}{f(r)} + r^2 d\theta^2 + r^2sin^2\theta d\phi^2
\end{equation}
where the lapse function $f(r) = 1-\frac{2M}{r}$. Determinant of the metric (\ref{B3}) is defined as
\begin{equation}\label{B4}
    g = det (g_{\mu\nu}) = - r^4 sin^2\theta
\end{equation}
The unit normal vector to a constant $r$ surface is defined as
\begin{equation}\label{B5}
    n^{\alpha}n_{\alpha} = 1 \Rightarrow n^r n_r = n^r g_{rr} n^r = 1 \Rightarrow n^r = \frac{1}{\sqrt{g_{rr}}} = \sqrt{f(r)}
\end{equation}
Now, the trace of the extrinsic curvature is defined as
\begin{multline}\label{B6}
    K = n^\alpha_{;\alpha} = \frac{1}{\sqrt{-g}}\frac{\partial}{\partial x^\alpha}[\sqrt{-g}\times n^\alpha]= \frac{1}{\sqrt{-g}}\frac{\partial}{\partial r}[\sqrt{-g}\times n^r]\\ = \frac{1}{r^2sin\theta}\frac{\partial}{\partial r}\left[r^2sin\theta\sqrt{f(r)}\right] = \frac{\left[r^2f'(r) + 4rf(r)\right]}{2r^2\sqrt{f(r)}}
\end{multline}
where $f'(r) = df(r)/dr$. The induced metric for a constant $r$ surface becomes
\begin{equation}\label{B7}
    ds^2 = -f(r)dt^2 + r^2d\theta^2 + r^2sin^2\theta d\phi^2
\end{equation}
Determinant of the metric (\ref{B7}) is defined as
\begin{equation}\label{B8}
    h = det (h_{ab}) = -r^4f(r)sin^2\theta
\end{equation}
Now, in the natural unit $\hbar = c = 1$ and $G = 1$, the Gibbon-Hawking-York boundary action is defined as
\begin{multline}\label{B9}
    \mathcal{A}_{GHY} = \frac{1}{8\pi G}\int_{\partial \mathcal{M}} \sqrt{|h|}K d^3x = \frac{1}{8\pi}\int \sqrt{|h|}K dt d\theta d\phi\\ = \frac{1}{8\pi}\int dt \int \frac{r^2\sqrt{f(r)}sin\theta}{2r^2\sqrt{f(r)}}\left[r^2f'(r) + 4rf(r)\right]d\theta d\phi
\end{multline}
The rate of change of the Gibbon-Hawking-York boundary action is defined as
\begin{multline}\label{B10}
    \frac{d\mathcal{A}_{GHY}}{dt} =  \frac{1}{16\pi}\left[r^2f'(r) + 4rf(r)\right]\int_{0}^{\pi} sin\theta d\theta \int_{0}^{2\pi} d\phi\\ =  \frac{1}{16\pi}\left[r^2f'(r) + 4rf(r)\right] \times 4\pi = \frac{1}{2}\left[2r-3M\right]
\end{multline}
The boundary of the Wheeler-DeWitt patch for a Schwarzschild black hole extends from the event horizon $r_+ = 2M$ to the singularity $r = 0$. Therefore, the GHY action at $r= 0$ and $r= r_+$ is
\begin{equation}\label{B11}
    \frac{d\mathcal{A}_{GHY}}{dt} = \frac{1}{2}\left[2r - 3M\right]_{0}^{2M} = 2M
\end{equation}
From Eqs. (\ref{B2}) and (\ref{B11}), the total rate of change of action becomes
\begin{equation}\label{B12}
    \frac{d\mathcal{A}}{dt} = \frac{d\mathcal{A}_{EH}}{dt} + \frac{d\mathcal{A}_{GHY}}{dt} = 2M
\end{equation}
Therefore, the complexity growth rate is defined as
\begin{equation}\label{B13}
    \frac{d\mathcal{C}}{dt} = \frac{1}{\pi \hbar}\frac{d\mathcal{A}}{dt} = \frac{2M}{\pi}
\end{equation}
where we put $\hbar = 1$ in the final expression. Now, the Hawking temperature and entropy are defined as
\begin{equation}\label{B14}
    T_H = \frac{M}{2\pi r^2_+}, \ \ S = \frac{A}{4} = \pi r^2_+\ \ \Rightarrow T_HS_H = \frac{M}{2}
\end{equation}
From Eqs. (\ref{B13}) and (\ref{B14}), the complexity growth rate in terms of product $T_HS_H$ becomes
\begin{equation}\label{B15}
    \frac{d\mathcal{C}}{dt} = \frac{4}{\pi} T_HS_H 
\end{equation}

\section{REISSNER-NORDSTR\"OM BLACK HOLE}\label{Appendix C}
\subsection{Einstein-Maxwell action}
The Einstein-Maxwell action is defined as
\begin{multline}\label{C1}
    \mathcal{A}_{EM} = -\frac{1}{16\pi}\int_{\mathcal{M}} \sqrt{-g}F_{\mu\nu}F^{\mu\nu}d^4x\\ = -\frac{1}{16\pi}\int \sqrt{-g}F_{\mu\nu}F^{\mu\nu}dt dr d\theta d\phi
\end{multline}
where the determinant of the metric is $g = -r^4sin^2\theta$ and nonzero components of the electric field strength are \cite{toolkit}
\begin{equation}\label{C2}
    F_{rt} = - F_{tr} = \frac{Q}{r^2}
\end{equation}
So the value of product $F_{\mu\nu}F^{\mu\nu}$ becomes
\begin{multline}\label{C3}
    F_{\mu\nu}F^{\mu\nu} = F_{rt}F^{rt} + F_{tr}F^{tr} = F_{rt}F^{rt} + (-F_{rt})(-F^{rt})\\ = 2F_{rt}F^{rt} = 2F_{rt} g^{rr} g^{tt} F_{rt} = 2F_{rt}\times(-1)\times F_{rt} = -\frac{2Q^2}{r^4}
\end{multline}
Now, from Eqs. (\ref{C1}) and (\ref{C3}), the rate of change of the action becomes 
\begin{multline}\label{C4}
    \frac{d\mathcal{A}_{EM}}{dt} =  -\frac{1}{16\pi}\int r^2sin\theta \times \left(\frac{-2Q^2}{r^4}\right)dr d\theta d\phi\\ = \frac{Q^2}{8\pi}\int_{r_-}^{r_+} \frac{dr}{r^2}\int_{0}^{\pi} sin\theta \int_{0}^{2\pi} d\phi = \frac{Q^2}{2} \left(\frac{1}{r_-} - \frac{1}{r_+}\right)\\ = \sqrt{M^2 - Q^2}
\end{multline}
\subsection{Gibbons-Hawking-York action}
The metric of a Reissner–Nordstr\"om black hole is defined as 
\begin{equation}\label{C5}
    ds^2= -f(r)dt^2 + \frac{dr^2}{f(r)} + r^2 d\theta^2 + r^2sin^2\theta d\phi^2
\end{equation}
where the lapse function $f(r) = 1-\frac{2M}{r} +\frac{Q^2}{r^2}$. Determinant of the metric (\ref{C5}) is defined as
\begin{equation}\label{C6}
    g = det (g_{\mu\nu}) = - r^4 sin^2\theta
\end{equation}
The unit normal vector to a constant $r$ surface is defined as
\begin{equation}\label{C7}
    n^{\alpha}n_{\alpha} = 1 \Rightarrow n^r n_r = n^r g_{rr} n^r = 1 \Rightarrow n^r = \frac{1}{\sqrt{g_{rr}}} = \sqrt{f(r)}
\end{equation}
Now, the trace of the extrinsic curvature is defined as
\begin{multline}\label{C8}
    K = n^\alpha_{;\alpha} = \frac{1}{\sqrt{-g}}\frac{\partial}{\partial x^\alpha}[\sqrt{-g}\times n^\alpha]= \frac{1}{\sqrt{-g}}\frac{\partial}{\partial r}[\sqrt{-g}\times n^r]\\ = \frac{1}{r^2sin\theta}\frac{\partial}{\partial r}\left[r^2sin\theta\sqrt{f(r)}\right] = \frac{\left[r^2f'(r) + 4rf(r)\right]}{2r^2\sqrt{f(r)}}
\end{multline}
where $f'(r) = df(r)/dr$. The induced metric for a constant $r$ surface becomes
\begin{equation}\label{C9}
    ds^2 = -f(r)dt^2 + r^2d\theta^2 + r^2sin^2\theta d\phi^2
\end{equation}
The determinant of the induced metric (\ref{C9}) is defined as
\begin{equation}\label{C10}
    h = det (h_{ab}) = -r^4f(r)sin^2\theta
\end{equation}
Now, in the natural unit $\hbar = c = 1$ and $G = 1$, the Gibbon-Hawking-York boundary action is defined as
\begin{multline}\label{C11}
    \mathcal{A}_{GHY} = \frac{1}{8\pi G}\int_{\partial\mathcal{M}} \sqrt{-h}K d^3x = \frac{1}{8\pi}\int\sqrt{-h}K dt d\theta d\phi\\ = \frac{1}{8\pi}\int dt \int \frac{r^2\sqrt{f(r)}sin\theta}{2r^2\sqrt{f(r)}}\left[r^2f'(r) + 4rf(r)\right]d\theta d\phi
\end{multline}
The rate of change of the Gibbons-Hawking-York boundary action is defined as
\begin{multline}\label{C12}
    \frac{d\mathcal{A}_{GHY}}{dt} =  \frac{1}{16\pi}\left[r^2f'(r) + 4rf(r)\right]\int_{0}^{\pi} sin\theta d\theta \int_{0}^{2\pi} d\phi\\ =  \frac{1}{16\pi}\left[r^2f'(r) + 4rf(r)\right] \times 4\pi = \frac{1}{4}\left[r^2f'(r) + 4rf(r)\right]
\end{multline}
At horizons where lapse function $f(r_\pm) = 0$, and with $f'(r) = df(r)/dr = \frac{2M}{r^2} - \frac{2Q^2}{r^3}$, the boundary of the Wheeler-DeWitt patch for a Reissner-Nordstr\"om black hole extends from the outer $r_+$ to the inner horizon $r_-$. Therefore, the GHY action at $r=r_-$ and $r= r_+$ is
\begin{multline}\label{C13}
    \frac{d\mathcal{A}_{GHY}}{dt} = \frac{1}{4}\left[r^2f'(r)\right]_{r_-}^{r_+} = \frac{1}{4}\left[2M -\frac{2Q^2}{r}\right]_{r_-}^{r_+}\\ = \frac{1}{2}\left[\frac{Q^2}{r_-} - \frac{Q^2}{r_+}\right] = \frac{Q^2}{2}\left(\frac{r_+ - r_-}{r_+ r_-}\right) = \sqrt{M^2 - Q^2}
\end{multline}
From Eqs. (\ref{C4}) and (\ref{C13}), the total rate of change of action becomes 
\begin{equation}\label{C14}
    \frac{d\mathcal{A}}{dt} = \frac{d\mathcal{A}_{EM}}{dt} + \frac{d\mathcal{A}_{GHY}}{dt} = 2\sqrt{M^2 - Q^2}
\end{equation}
Hence, the complexity growth rate becomes
\begin{equation}\label{C15}
    \frac{d\mathcal{C}}{dt} = \frac{1}{\pi \hbar}\frac{d\mathcal{A}}{dt} = \frac{2}{\pi}\sqrt{M^2 - Q^2}
\end{equation}
where we substitute $\hbar = 1$ in the final expression. Now, the Hawking temperature and entropy are defined as
\begin{multline}\label{C16}
  T_H = \frac{1}{2\pi}\frac{\sqrt{M^2 - Q^2}}{ (r_+^2 + Q^2)}, \ \ S_H = \frac{A}{4}= \pi (r_+^2 + Q^2)\\
  \Rightarrow T_H S_H = \frac{1}{2}\sqrt{M^2 - Q^2}
\end{multline}
From Eqs. (\ref{C15}) and (\ref{C16}), the complexity growth rate in terms of product $T_HS_H$ becomes
\begin{equation}\label{C17}
    \frac{d\mathcal{C}}{dt} = \frac{4}{\pi}T_H S_H 
\end{equation}
\section{KERR BLACK HOLE}\label{Appendix D}
\subsection{Einstein-Hilbert action}
The Einstein-Hilbert bulk action is defined as
\begin{equation}\label{D1}
   \mathcal{A}_{EH} = \frac{1}{16\pi G}\int \sqrt{-g} R d^4x = \frac{1}{16\pi G}\int \sqrt{-g} R dt dr d\theta d\phi
\end{equation}
We know the Ricci scalar for the Kerr black hole in the vacuum space $R = 0$. Hence, the rate of change of action is
\begin{equation}\label{D2}
    \frac{d \mathcal{A}_{EH}}{dt} = \frac{1}{16\pi }\int \sqrt{-g} R dr d\theta d\phi = 0
\end{equation}
\subsection{Gibbons-Hawking-York action}
The Kerr metric in the Boyer-Lindquist coordinates is defined as
\begin{multline}\label{D3}
    ds^2 = - \frac{(\Delta - a^2sin^2{\theta})}{\rho^2}dt^2- \frac{4Mra sin^2{\theta}}{\rho^2} dtd\phi + \frac{\rho^2}{\Delta}dr^2\\ + \rho^2d{\theta^2} + \frac{Asin^2{\theta}}{\rho^2}d\phi^2 
\end{multline}
The parameters, $\Delta, \rho^2, a,$ and $A$ are defined as
\begin{equation}\label{D4}
    \begin{split}
    \Delta = r^2 - 2Mr + a^2,\ \ \rho^2 = r^2 + a^2cos^2{\theta}\\ 
   a = J/Mc, \ \ A  = (r^2 + a^2)^2 - \Delta a^2sin^2{\theta}
    \end{split}
\end{equation}
Here, $M$ and $J$ are the ADM mass and angular momentum of the Kerr black hole. Determinant of metric (\ref{D3}) is defined as
\begin{equation}\label{D5}
    g = det(g_{\mu\nu}) = -\rho^4sin^2\theta
\end{equation}
The unit normal vector to a constant $r$ surface is defined as
\begin{equation}\label{D6}
    n^{\alpha}n_{\alpha} = 1 \Rightarrow n^r n_r = n^r g_{rr} n^r = 1 \Rightarrow n^r = \frac{1}{\sqrt{g_{rr}}} = \sqrt{\frac{\Delta}{\rho^2}}
\end{equation}
Now, the trace of the extrinsic curvature is defined as
\begin{multline}\label{D7}
    K = n^\alpha_{;\alpha} = \frac{1}{\sqrt{-g}}\frac{\partial}{\partial x^\alpha}[\sqrt{-g}\times n^\alpha]= \frac{1}{\sqrt{-g}}\frac{\partial}{\partial r}[\sqrt{-g}\times n^r]\\ = \frac{1}{\rho^2sin\theta}\frac{\partial}{\partial r}\left[\rho^2sin\theta\sqrt{\frac{\Delta}{\rho^2}}\right] = \frac{1}{\rho^2}\frac{\partial}{\partial r}\left[\sqrt{\rho^2\Delta}\right]\\ = \frac{1}{2\rho^2\sqrt{\rho^2\Delta}}[2r\Delta + 2(r-M)\rho^2]
\end{multline}
The induced metric for a constant $r$ surface becomes
\begin{multline}\label{D8}
    ds^2 = - \frac{(\Delta - a^2sin^2{\theta})}{\rho^2}dt^2- \frac{4Mra sin^2{\theta}}{\rho^2} dtd\phi + \rho^2d{\theta^2}\\ + \frac{Asin^2{\theta}}{\rho^2}d\phi^2 
\end{multline}
Determinant of the induced metric (\ref{D8}) is defined as
\begin{equation}\label{D9}
    h = det (h_{ab}) = -\rho^2\Delta sin^2\theta
\end{equation}
Now, in the natural unit $\hbar = c = 1$ and $G = 1$, the Gibbon-Hawking-York boundary action is defined as
\begin{multline}\label{D10}
     \mathcal{A}_{GHY} = \frac{1}{8\pi G}\int_{\partial\mathcal{M}} \sqrt{|h|}K d^3x = \frac{1}{8\pi}\int \sqrt{|h|}K dt d\theta d\phi\\ = \frac{1}{8\pi}\int dt \int \sqrt{\rho^2\Delta}sin\theta \times\frac{[2r\Delta + 2(r-M)\rho^2]}{2\rho^2\sqrt{\rho^2\Delta}} d\theta d\phi \\
     = \frac{1}{8\pi}\int dt \int_{0}^{\pi}\left[\frac{rsin\theta\Delta}{\rho^2} + (r-M)sin\theta\right]d\theta\int_{0}^{2\pi} d\phi\\ = \frac{1}{4}\int dt \int_{0}^{\pi}\left[\frac{rsin\theta\Delta}{\rho^2} + (r-M)sin\theta\right]d\theta
\end{multline}
At horizons where $\Delta(r_\pm) = 0$. The boundary of the Wheeler-DeWitt patch for a Kerr black hole extends from the outer $r_+$ to the inner horizon $r_-$. Therefore, the GHY action at $r=r_-$ and $r= r_+$ is
\begin{multline}\label{D11}
    \frac{d\mathcal{A}_{GHY}}{dt} = \frac{1}{4}\left[r-M\right]_{r_-}^{r_+}\int sin\theta d\theta = \frac{1}{2}\left[r-M\right]_{r_-}^{r_+}\\ = \frac{1}{2}\left[2\sqrt{M^2-a^2}\right] = \sqrt{M^2-a^2}
\end{multline}
From Eqs. (\ref{D2}) and (\ref{D11}), the total rate of change of action is
\begin{equation}\label{D12}
    \frac{d \mathcal{A}}{dt} = \frac{d \mathcal{A}_{EH}}{dt} + \frac{d\mathcal{A}_{GHY}}{dt} = \sqrt{M^2-a^2}
\end{equation}
Hence, the complexity growth rate becomes
\begin{equation}\label{D13}
    \frac{d\mathcal{C}}{dt} = \frac{1}{\pi \hbar}\frac{d\mathcal{A}}{dt} = \frac{1}{\pi}\sqrt{M^2 - a^2}
\end{equation}
where we substitute $\hbar = 1$ in the final expression. Now, the Hawking temperature and entropy are defined as
\begin{multline}\label{D14}
  T_H = \frac{1}{2\pi}\frac{\sqrt{M^2 - a^2}}{ (r_+^2 + a^2)}, \ \ S_H = \frac{A}{4}= \pi (r_+^2 + a^2)\\
  \Rightarrow T_H S_H = \frac{1}{2}\sqrt{M^2 - a^2}
\end{multline}
From Eqs. (\ref{D13}) and (\ref{D14}), the complexity growth rate in terms of product $T_HS_H$ becomes
\begin{equation}\label{D15}
    \frac{d\mathcal{C}}{dt} = \frac{2}{\pi}T_H S_H
\end{equation}
\section{VARIATION IN THE COMPLEXITY GROWTH RATE FOR KERR BLACK HOLE}\label{Appendix E}
The CV conjecture connects the complexity with the volume of black holes, and hence we expect that the variation in the complexity growth rate $\delta\mathcal{\dot C}\ (\mathcal{\dot C}= {d\mathcal{C}/dt})$ might be directly related to $\delta\mathcal{\dot V} $. From Eq. (\ref{eqn9}), the $\delta \mathcal{\dot C}$ for Kerr black hole becomes
\begin{equation}\label{E2}
    \delta \left[\frac{d\mathcal{C}}{dt}\right] \sim \delta \left[\frac{1}{r_+}\frac{dV}{dt}\right]\ \ \Rightarrow \delta\mathcal{\dot C}\sim \frac{1}{r^2_+}\left[r_+\delta\mathcal{\dot V} - \mathcal{\dot V}\delta r_+ \right]
\end{equation}
where $\delta \mathcal{\dot{V}}$ is the variation in the volume rate and $\delta r_+$ is the variation in the horizon radius. In our earlier work \cite{SSR3}, we have shown the variation in the volume rate $ \delta\mathcal{\dot V}$ as
\begin{equation}\label{E3}
   \begin{split}
        \delta \mathcal{\dot{V}} = 6\sqrt{3}\pi M\bigg[(\delta M - \Omega_H\delta J) - \frac{37}{27}\Omega_H\delta J\bigg]\\
        \Rightarrow (\delta M - \Omega_H\delta J) = \frac{\delta \mathcal{\dot{V}}}{6\sqrt{3}\pi M} + \frac{37}{27}\Omega_H\delta J
   \end{split}
\end{equation}\\
\textbf{Calculation of $\delta r_+$ :} The event horizon of the Kerr black hole is defined as $r_+ = M + \sqrt{M^2 -a^2}$, so the variation $\delta r_+$ in a small $a/M$ limit becomes
\begin{equation}\label{E4}
    \begin{split}
        r_+ = M + \sqrt{M^2 -a^2} = M + \sqrt{M^2 - (J/M)^2}\\
        \delta r_+ = \delta M + \frac{[2M\delta M + \frac{2J^2\delta M}{M^3}- \frac{2J\delta J}{M^2}]}{2\sqrt{M^2 - (J/M)^2}}
    \end{split}
\end{equation}
In the limit $J\ll M$, we can neglect higher power terms of $J$, and $\sqrt{M^2 - (J/M)^2} \approx M$, so we get
\begin{equation}\label{E5}
     \delta r_+ =  \delta M + \frac{1}{2M}\left[2M\delta M - \frac{2J\delta J}{M^2}\right]  = 2\delta M -  \frac{J}{M^3}\delta J
\end{equation}
The horizon's angular momentum $\Omega_H$ in a small $a/M$ limit is defined as
\begin{equation}\label{E6}
    \Omega_H = \frac{a}{r^2_+ + a^2} = \frac{J/M}{2M^2 + 2M^2\sqrt{1 - J^2/M^4}} \approx \frac{J}{4M^3}
\end{equation}
From Eqs. (\ref{E5}) and (\ref{E6}), we get
\begin{equation}\label{E7}
    \delta r_+ = \left[2(\delta M - \Omega_H \delta J)-2\Omega_H \delta J\right] 
\end{equation}
From Eqs. (\ref{E3}) and (\ref{E7}), we get
\begin{multline}\label{E8}
    \delta r_+ = 2\bigg(\frac{\delta \mathcal{\dot{V}}}{6\sqrt{3}\pi M} + \frac{37}{27}\Omega_H\delta J\bigg)-2\Omega_H \delta J \\
    = \frac{\delta \mathcal{\dot{V}}}{3\sqrt{3}\pi M} + \frac{74}{27}\Omega_H\delta J - 4\Omega_H \delta J\\
    = \frac{\delta \mathcal{\dot{V}}}{3\sqrt{3}\pi M} - \frac{34}{27}\Omega_H\delta J = \frac{0.061}{M}\delta \mathcal{\dot{V}} - 1.26\Omega_H\delta J 
\end{multline}
Hence, the product of volume rate $\mathcal{\dot{V}}$ and $\delta r_+$ gives
\begin{equation}\label{E9}
    \mathcal{\dot{V}}\delta r_+ = 0.061\frac{\mathcal{\dot{V}}}{M}\delta \mathcal{\dot{V}} - 1.26\mathcal{\dot{V}}\Omega_H\delta J
\end{equation}\\
\textbf{Calculation of $\delta \mathcal{\dot C}$ :} Substituting the value of $\mathcal{\dot{V}}\delta r_+$, from Eq. (\ref{E9}) into Eq. (\ref{E2}), we get
\begin{multline}\label{E10}
      \delta \mathcal{\dot C} \sim \frac{1}{r^2_+}\bigg[r_+\delta\mathcal{\dot V} -  0.061\frac{\mathcal{\dot{V}}}{M}\delta \mathcal{\dot{V}} + 1.26\mathcal{\dot{V}}\Omega_H\delta J\bigg]\\ 
      \sim \frac{1}{r^2_+}\bigg[\bigg
      (r_+ -  0.061\frac{\mathcal{\dot{V}}}{M}\bigg)\delta \mathcal{\dot{V}} + 1.26\mathcal{\dot{V}}\Omega_H\delta J\bigg]
\end{multline}
If in the limit when $J\ll M$, $r_+\rightarrow 2M, \mathcal{\dot V}/M\rightarrow 16.32M$, and $r_+ - 0.061\mathcal{\dot V}/M\rightarrow 1.0045M$. Substituting these values in the above equation, we get
\begin{multline}\label{E11}
     \delta \mathcal{\dot C} \sim \frac{1}{4M^2}\left[\left
      (1.0045\right)M\delta \mathcal{\dot{V}} + 20.56M^2\Omega_H\delta J\right]\\ \sim \frac{1}{4M}\left[\left
      (1.0045\right)\delta \mathcal{\dot{V}} + 20.56M\Omega_H\delta J\right]
\end{multline}
Substituting the value of $\delta \mathcal{\dot V}$ from Eq. (\ref{E3}) into Eq. (\ref{E11}), we get
\begin{multline}\label{E12}
    \delta \mathcal{\dot C} \sim \frac{1}{4M}\bigg[6\sqrt{3}\pi M\bigg\{(\delta M - \Omega_H\delta J) - \frac{37}{27}\Omega_H\delta J\bigg\}(1.0045)\\ + 20.56M\Omega_H\delta J\bigg]\\
     \sim \frac{1}{4}\left[32.795 (\delta M - \Omega_H\delta J) - 44.941\Omega_H\delta J + 20.56\Omega_H\delta J\right]\\
     \sim \frac{1}{4}\left[32.795 (\delta M - \Omega_H\delta J) - 24.38\Omega_H\delta J \right]\\
     \sim 8.198 (\delta M - \Omega_H\delta J) - 6.095\Omega_H\delta J\\
     \sim 8.198\left[ (\delta M - \Omega_H\delta J) - 0.74\Omega_H\delta J\right]
\end{multline}
Hence, the variation in the complexity growth rate approximately becomes
\begin{equation}\label{E13}
    \delta \mathcal{\dot C} \sim 8\left[ (\delta M - \Omega_H\delta J) - 0.74\Omega_H\delta J\right]
\end{equation}
This is an important expression that demonstrates interesting behavior under various physical processes, including the Penrose process, superradiance, particle accretion, and Hawking radiation.
\bibliography{apssamp}

\end{document}